\documentclass{aptpub}

\usepackage{amsmath}
\usepackage{graphicx}
\usepackage{amssymb}
\usepackage{subcaption}
\newcommand{\beqn}{\begin{eqnarray*}}
\newcommand{\eeqn}{\end{eqnarray*}}
\newcommand{\bneqn}{\begin{eqnarray}}
\newcommand{\eneqn}{\end{eqnarray}}
\newcommand{\parens}[1]{\left(#1\right)}
\newcommand{\tothepow}[2]{\parens{#1}^{#2}}
\newcommand{\bracks}[1]{\left[#1\right]}
\newcommand{\expe}[1]{\mathbb{E}\bracks{#1}}
\newcommand{\expesub}[2]{\mathbb{E}_{#1}\bracks{#2}}
\newcommand{\cexpe}[2]{\expe{#1 \, | \, #2}}
\newcommand{\sgn}[1]{\text{sgn}\parens{#1}}
\newcommand{\expesubsup}[3]{\mathbb{E}_{#1}^{#2}\bracks{#3}}
\renewcommand{\exp}[1]{\mathrm{exp}\parens{#1}}

\authornames{Philip A. Ernst, Dean P. Foster, Larry A. Shepp}
\shorttitle{How to Retire Early}

\begin{document}
\title{On Optimal Retirement}

\authorone[The Wharton School, University of Pennsylvania] {Philip A. Ernst}
\addressone{3730 Walnut Street, Philadelphia, PA 19104}

\authortwo[The Wharton School, University of Pennsylvania] {Dean P. Foster}
\addresstwo{3730 Walnut Street, Philadelphia, PA 19104}

\authorthree[The Wharton School, University of Pennsylvania] {Larry A. Shepp}
\addressthree{3730 Walnut Street, Philadelphia, PA 19104}

\begin{abstract}

We pose an optimal control problem arising in a perhaps new model for retirement investing. Given a control function $f$ and our current net worth as $X(t)$ for any $t$, we invest an amount $f(X(t))$ in the market. We need a fortune of $M$ ``superdollars" to retire and want to retire as early as possible. We model our change in net worth over each infinitesimal time interval by the Ito process $dX(t)= (1+f(X(t))dt+ f(X(t))dW(t)$. We show how to choose the optimal $f=f_0$ and show that the choice of $f_0$ is optimal among all nonanticipative investment strategies, not just among Markovian ones.

\end{abstract}


\section{Introduction}\label{sec:introduction}

We begin by discussing the rationale and assumptions underlying the process chosen to model our change in wealth over each infinitesimal time interval. The model employs deflated (``constant") dollars and assumes that the investor borrows at the risk-free rate. The optimal control function ensures that the investor will only be able to borrow an amount of money such that, with probability 1, $X(t) \ge 0$ for all $t$. Like \cite{Cover}, the model assumes that the investor can either put money into or take out money out of the market in continous time and that there is no transaction fee for doing so. 

The investor chooses to invest solely in the ``Sharpe asset," explicitly defined on page 12 of \cite{Foster}.  \cite{Foster} proves that the Sharpe asset has close to unit variance (this is why $\sigma=1$ in the volatility term $\sigma f(X(t))dW(t)$). In formalizing the Ito process, we first normalize our unit of time so that the Sharpe asset is expected to return 100 percent over a unit time interval. This normalization forces $r=1$ in the term $rf(X(t)dt$. We proceed to normalize salary. Our monetary unit, which we will call a ``superdollar," is normalized such that we have a steady income of $dt$ in each time interval of length $dt$. We need a fortune of $M$ superdollars to retire and wish to retire as early as possible. See \cite{Merton2} and \cite{Merton} for broader economic discussion of retirement processes.

The solution to this optimal control problem in retirement investing is especially interesting because it involves what seems to be a new phenomenon in boundary behavior, where a process hits a boundary and
reflects ``softly'' from it, without the need for local time as in reflecting
Brownian motion. We show the reflection takes place by using what seems to be
a new approach to stochastic differential equations which avoids the clock
changing methods of \cite{IandM} and \cite{HPM}, namely by first defining a particular
``unit'' diffusion, which is a diffusion where the diffusion coefficient is
identically unity, which is then easier to construct using Picard's method than
a general diffusion. If we then take a particular monotonic function of the
unit diffusion to construct the diffusion which is obtained as the answer,
$f_0$, above, where the monotone function is everywhere nonnegative, it follows
that the final diffusion is also always nonnegative.

We also consider a more general model of retirement investing where the
diffusion coefficient, $f$, is replaced by $Af^\alpha$. The general case
reveals that the optimal investment strategy as well as the expected time
until retirement are very strongly dependent in interesting 
ways on the particular model used.

\section{Formal Model and Approach}\label{sec:formal_model_approach}

Formally, the problem is stated as follows: given $0 \le x \le M$, and a Brownian motion, $W(t), ~t \ge 0$,
to find a nonanticipating process, $f(t), ~t \ge 0$, so that if $\tau^f_M$ is
the hitting time of $M$ of the Ito process $X(t) = X^f(t), ~t \ge 0$, with
$X(0) = x$, and
\beqn
dX(t) = (1+f(X(t)))dt + f(X(t)) dW(t)
\eeqn
then $V(x;f) = \expesub{x}{\tau^f_M}$ is a minimum over all such allowable $f$. Note that
we are assuming that the state space for this optimal control problem is the 
right half line so that we do not allow negative values of $X(t)$. Later, we will prove that, in our setup, $X(t)$ cannot be negative. If
$X(t) < 0$ for some $t < \tau^f_M$, which is possible if $f(t)$ is bounded
away from zero and also bounded, then there is a need to define what happens
if the investor is in debt; we assume the game is over in this case and then
$\tau^f_M = \infty$ so that with our definition, we do not even achieve a
finite expectation, much less a minimum. Other definitions allow for borrowing additional capital, but our model
assumes that we are extremely adverse to being in debt.

We will show that under this assumption, the optimum control, $f_0$, exists and
is unique. Any reasonable person would guess that $f(t) = f(X(t))$, i.e., 
that the optimal $f$ is ``Markovian'', i.e., the optimal strategy depends only
on the present fortune. But even if we guess that $f$ should be Markovian, how
do we learn which particular $f$ is best? There is a nice way, involving a lot
of nice guessing. Once one guesses $f$ the proof that it is optimal is routine
crank-turning, by martingale theory as we will see.

\section{Formal Statement of Results}\label{sec:formal_statement_results}

To get lower bounds on $V(x) = \inf_f V(x;f)$, one needs to find, in the usual
way, a function, ${\bar{V}}(x)$, with ${\bar{V}}(M) = 0$, for which, for any
$f$, the process, $Y(t) = t + {\bar{V}}(X^f(t) )$ is a submartingale. If this
is the case, then, we have from optional sampling $\expesub{x}{Y\parens{\tau^f_M}} \ge Y(0)$.
This gives that for any $f$ and $0 \le x \le M$,
\beqn
\expesub{x}{\tau^f_M} = \expe{Y(\tau^f_M)} \ge Y(0) = {\bar{V}}(x)
\eeqn
and since this holds for any $f$ and $0 \le x \le M$, we get that $V(x) \ge {\bar{V}}(x)$.

Equality will hold for all $x$, for the greatest lower bound, ${\bar{V}}$.
The class of all such ${\bar{V}}$'s is a convex class determined by the
Ito inequalities defining a submartingale, which are that ${\bar{V}} \ge 0$,
and that for all $x$ and all $f$,
\beqn
\cexpe{dY(t)}{\mathcal{F}_t} = {\bar{V}}^\prime(x) (1 + f) dt + \frac{f^2}{2}{\bar{V}}^{\prime \prime}(x) dt + dt \ge 0
\eeqn

Since this must hold for all choices of $f$ and all choices of $x = X(t)$ in
$[0,\infty)$, and since this is quadratic in the real variable $f$ (if this
seems somewhat aggressive with respect to logic, recall that we are just using
this reasoning for \textit{guessing} the right ${\bar{V}}$). For any such
${\bar{V}}$, we have that for any $f$, $V(x;f) \ge {\bar{V}}(x)$, which gives
us the lower bound, ${\bar{V}}(x)$ on $V(x)$. Which ${\bar{V}}$'s satisfy the
above submartingale condition?

Setting the derivative wrt. $f$ equal to zero we see that we must have for each
$x$, ${\bar{V}}^{\prime \prime}(x) > 0$, and then the minimum occurs at
$f = f(x) = -\bar{V}^\prime(x) / \bar{V}^{\prime \prime}(x)$.
Putting this $f$ back into the submartingale inequality we need that
\beqn
1 + {\bar{V}}^\prime(x) - \frac{1}{2}\frac{({\bar{V}}^\prime)^2(x)}{{\bar{V}}^{\prime \prime}(x)} \ge 0
\eeqn

For the best $f$, we need equality to hold everywhere in the string of
inequalities above so that we would choose ${\bar{V}}$ to satisfy the last
inequality with equality throughout. If we set 
\beqn
g(x) = -{\bar{V}}^\prime(x) \ge 0
\eeqn

then we seek $g$ to satisfy
\beqn
g^\prime(x) \parens{\frac{1}{g(x)} - \frac{1}{g^2(x)}} \equiv \frac{1}{2}
\eeqn

Integrating, we have for some integration constant $c$
\beqn
\frac{1}{g(x)} + \log{(g(x))} = \frac{x+c}{2}
\eeqn

Since the left side is of the form $\frac{1}{y} + \log{y} \ge 1$ for all 
$y > 0$, it is tempting to choose $c = 2$ since this makes the right side
greater than or equal to one for $x \ge 0$ (it is our privilege to do
this, since we are just guessing). We have almost arrived at a guess for the 
best ${\bar{V}}$, namely we have to solve the last equation for
$g(x) = -{\bar{V}}^\prime(x)$, and then ${\bar{V}}$ is determined because we
have ${\bar{V}}(M) = 0$.

A plot of $y$ vs. $\frac{1}{y} + \log{y}$ is given in Figure 1 which shows
that the inverse function defining $g(x)$ by
\beqn
\frac{1}{g(x)} + \log{(g(x))} = 1+ \frac{x}{2}
\eeqn

is not unique since the inverse is not one-one. 

\begin{figure}[htp]
\centering
\includegraphics[width=2.5in]{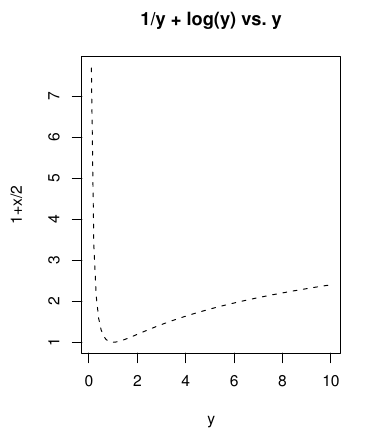}
\caption{$y$ vs. $\frac{1}{y} + \log{y}$ }
\label{fig:fig1}
\end{figure}

Which one do we use, the left
side branch or the right side branch to define $g(x)$ for each $x \in [0,M]$?
Recall that we must have $g^\prime(x) = -{\bar{V}}^{\prime \prime}(x) < 0$, so
we guess to use the left side to determine $g(x)$. There is then clearly a
unique solution, $g(x)$, and we declare this as our guess at $g(x) = -{\bar{V}}^\prime(x)$.

Using the condition ${\bar{V}}(M) = 0$, we have 
\beqn
{\bar{V}}(x) = -\int_x^M{\bar{V}}^\prime(u) du = \int_x^M g(u) du
\eeqn

We have already set up the proof that this ${\bar{V}}(x) \equiv V(x)$.
We have also seen that any optimal choice of $f = f_0$ must satisfy $f(x) = -{\bar{V}}^\prime(x)/{\bar{V}}^{\prime \prime}(x)$, which we can express in terms of the $g(x)$ we have already defined because we
have seen that $V^\prime$ can be expressed in terms of $g$, and so we get that
\beqn
f_0(x) = \frac{g(x)}{-g^\prime(x)} = 2\parens{\frac{1-g(x)}{g(x)}}
\eeqn

Near $x = 0$, we can easily show through Maclaurin expansions for $\dfrac{1}{1-\sqrt{x}}$ and $\text{log}\parens{1-\sqrt{x}}$ that $g(x) \sim 1 - \sqrt{x}$, and so it
follows that 

\beqn
f_0(x) = 2\parens{\frac{1}{g(x)} -1} \sim 2 \sqrt{x}
\eeqn
For an approximation of $g(x)$ when $x$ is large, recall that we have defined $g(x)$ for $x\geq 0$ as the smaller of the two
solutions of:

\bneqn \label{eq:equation1}
\frac{1}{g(x)} + \log{(g(x))} = 1+ \frac{x}{2}
\eneqn

We proceed to define:

\beqn
a=\frac{1}{g(x)}\,\,\,\,\,\, b=1+\frac{x}{2}
\eeqn

Rewriting (\ref{eq:equation1}) in terms of $a$ and $b$, we arrive at:

\bneqn \label{eq:equation2}
a-\text{log}\, a=b
\eneqn

We now need an approximation for $a$ in terms of $b$. \\

We guess:

\bneqn \label{eq:equation3}
a \approx b + \text{log}\, b
\eneqn

Substituting our approximation into (\ref{eq:equation2}) we obtain:

\beqn
b + \text{log} \,b-\text{log}(b + \text{log}\,b)= b-\text{log}\parens{1+\frac{\text{log}\,b}{b}}
\eeqn

When $b$ is large, $\frac{\text{log}\,b}{b}$ is small, so the left hand side of (\ref{eq:equation2})  is very close to $b$. Substituting our definitions of $a$ and $b$ into (\ref{eq:equation3}), we obtain:

\beqn
\frac{1}{g(x)} \approx 1+ \frac{x}{2}+\text{log}\parens{1+\frac{x}{2}}
\eeqn

Or, equivalently,

\beqn
g(x) \approx \frac{1}{1+ \frac{x}{2}+\text{log}\parens{1+\frac{x}{2}}}
\eeqn

\beqn
f_0(x)= 2\parens{\frac{1}{g(x)}-1}\approx x+2 \, \text{log}\parens{1+\frac{x}{2}}
\eeqn

To complete that proof that this $f = f_0$ is optimal, with this choice of
$f = f_0$, and ${\bar{V}}$, the inequalities now hold for every other choice
of $f$ that $Y^f(t) = t+{\bar{V}}^f(X^f(t))$ is a submartingale. It follows
that $V(x) \ge {\bar{V}}(x)$ for all $x > 0$. We need to show that equality
holds for $f = f_0$ given above. It is true that in this case the submartingale
is a {\em local} martingale, except possibly at zero. We need to show the
equality $\expe{Y(\tau_M)} = Y(0)$ holds, where $Y$ is the process,
$Y(t) = t + {\bar{V}}(X^f_0(t) )$. It is enough to prove that $\tau_M < \infty$
w.p. 1. The difficulty is that the process, $X(t) = X^{f_0}(t)$ hits zero
uncountably many times with positive probability starting from any
$0 \le x < M$. How do we know that $X$ cannot take negative values? When
$X(t) = 0$, then the unit drift moves it to the right, but how do we know that
the term $f_0(X(t) ) dW(t)$ does not cause the process to reach the negative
half-line, or to get stuck at zero? Intuitively, each time the process hits
zero, imagine that $f(x)$ is turned off, so that $f(x) = 0$ for
$0 \le x \le \epsilon$. There is still a unit drift present so that the process
takes time $\epsilon$ to reach the point $x = \epsilon$. The probability
starting at $\epsilon$ that the process hits $M$ before it reaches zero again
is easily seen to be $1- c\epsilon$. It seems to follow from this that the
expected time to reach $M$ starting from any $x$ is finite and the conclusion
seems to follow. However, there is a mystery as to how the process pushes off
from zero and we shall now resolve this point.

It seems remarkable that the process $X^{f_0}$ behaves as if there is a
reflecting barrier at zero. It does not pass through zero to the negative
half-axis because $f_0(0) = 0$. This means that it slows down as it gets
near zero, but, unlike the Black-Scholes process \cite{Scholes}, it actually hits zero.
The drift, $1+f(0) = 1$, so that it then moves away from zero but it hits zero
uncountably many times (if it hits it once), just as the reflecting Brownian
process, $|W(t)|$, does, because the set of zeros of $X(t)$ is a perfect set.

It is instructive to consider a closely related reflecting process, the process
\beqn
dX(t) = dt + 2\sqrt{X(t)} dW(t)
\eeqn

How does one prove that $X(t) \ge 0$ for all $t$? This appears to be difficult
because $X(t) = 0$ uncountably many times, until one realizes that $X(t)$ is
simply $W^2(t)$. 
\begin{thm}
$X$ reflects off zero even though there is no local time in its Ito representation. 
\end{thm}

\begin{proof}
It is interesting that $X$ reflects off zero even though there is no local
time in its Ito representation. The way we prove that this makes sense seems to
be new and it seems to shed much light on the basic existence proofs of Ito
theory. In Ito theory, Picard iteration shows easily that a diffusion, 
$X = X(t,\omega)$ with a {\em unit} diffusion coefficient,
\beqn
dX(t) = A(X(t) )dt + dW(t), ~~t \ge 0, ~~X(0) = x_0
\eeqn
can be constructed on any space on which a process $W(t,\omega)$ with
continuous sample paths is available, path by path, if $A$ has bounded
difference quotients. It is then possible to construct a very general
diffusion, $Y(t)$, satisfying
\beqn
dY(t) = a(Y(t) ) dt + b(Y(t) ) dW(t)
\eeqn
by constructing the unit diffusion, $X(t)$, with appropriate drift, $A(x)$,
and then setting $Y(t) = g(X(t) )$, for an appropriate function, $g = g(x)$.
It is easy to check that to make $Y$ into an $a(y),b(y)$ diffusion, we must
choose $A(x), g(x)$ as follows:
\beqn
g^\prime(x) = b(g(x)) \quad \text{and} \quad A(x) = \frac{a(g(x)) -\frac{1}{2} b^\prime(g(x))g^\prime(x)}{b(g(x))}
\eeqn

If we do this for the case of $f_0$ which solves the problem, we see that
$g(x) \ge 0$, so it follows that the diffusion never goes negative and
reflects softly at $x = 0$, just as in the case of $W^2(t), ~t \ge 0$.

It seems remarkable that $g$ is defined only by the diffusion coefficient, $b$.
But this is somewhat illusory, because $A$ enters as well in that the unit
diffusion process, $X$, depends on $A$, and if $A$ does not satisfy a Lipschitz
condition then the domain of the $X$ diffusion may be a subset of the whole
line. Let us look at the case
\beqn
dY(t) = dt + c\sqrt{|Y(t)|} dW(t), ~~t \ge 0, ~~Y(0) = 0
\eeqn
in more detail since
this is an example which is very similar to the optimal control process of the
paper. We can carry out the steps of the determination of $B,g,A$, above to see
that
\beqn
B(y) = \int_0^y \frac{du}{c\sqrt{u}} = \frac{2}{c}\sqrt{|y|} \sgn{y}, \quad g(x) = \frac{c^2x^2}{4} \sgn{x}, \quad A(x) = \frac{2\parens{1-\frac{c^2}{4} \sgn{x}}}{c^2|x|}
\eeqn

We see there is an infinite singularity in $A$ at $x = 0$ except if $c = 2$.
If $0 < c < 2$ the process $X$ never hits zero, but if $c > 2$, then $X(t)$
cannot be defined after it hits zero, at least not by the diffusion equation
above.
\end{proof}

This completes the proof that for the optimal investment strategy, $f = f_0$,
the fortune of the young man reaches $M$ in a finite time with minimum expected
value. It is remarkable that the young man goes broke repeatedly with positive
probability before achieving his goal.

\begin{figure}[htp]
        \centering
        \begin{subfigure}[c]{0.4\textwidth}
                \centering
                \includegraphics[width=1.5in]{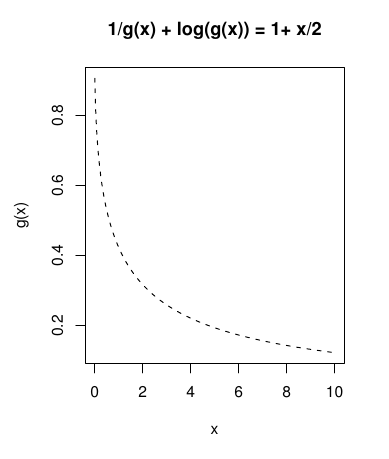}
                \caption{}
                \label{fig:fig2}
        \end{subfigure}%
        ~ 
        \begin{subfigure}[c]{0.4\textwidth}
                \centering
                \includegraphics[width=1.5in]{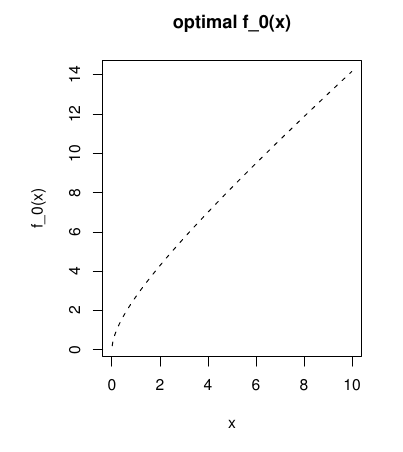}
                \caption{}
                \label{fig:fig3}
        \end{subfigure}\\
        ~ 
        \begin{subfigure}[c]{0.4\textwidth}
                \centering
                \includegraphics[width=1.5in]{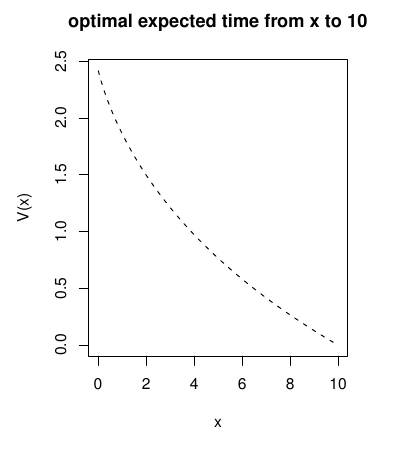}
                \caption{}
                \label{fig:fig4}
        \end{subfigure}
        \caption{Left: Plot of $g = g(x); 1/g + log(g) = 1 + x/2$, Right: Plot of the optimal $f = f_0, f (x) = \dfrac{g(x)}{g'(x)}$, Bottom: Plot of the optimal payoff, $V_{M}^{f}(x), M = 10, f = f_0$ }
\end{figure}

\begin{remark}
It is often remarked of some rich people that because they were
``aggressive, they went into bankruptcy several times before making it."
Somehow, mathematics seems to have already been aware of this common
observation! Note that it is always true that $V(x) \le M-x$ since an investor
can always choose $f \equiv 0$ and ``save the way to retirement''.
\end{remark}
A graph of $g(x) = - V^\prime(x)$ is given in Figure \ref{fig:fig2}, a graph of the optimal
payoff, $V(x)$, is given in Figure \ref{fig:fig3}, and a graph of the optimal investment
strategy, $f_0$, is given in Figure \ref{fig:fig4}. 

\section{Generalization of the problem}

A more general model for retirement than the one of Section \ref{sec:formal_statement_results}, namely,
\beqn
dX(t) = (1+f(X(t))) dt + f(X(t) ) dW(t), ~~ X(0) = x
\eeqn
would allow the diffusion term to be any fixed function of $f(X(t) )$ rather
than simply $f(X(t) )$ itself. The most natural choice was made above because
this is used in the Black-Scholes-Samuelson model for stock prices which was
arrived at under the argument that doubling an investment empirically seems to
double the volatility, but this is a crude argument and other possibilities
seem to be worth exploring. We propose considering the more general model:
\beqn
dX(t) = (1+f(X(t) ) dt + \phi(f(X(t))) dW(t), ~~ X(0) = x
\eeqn
where $\phi(u)$ is any increasing function. For tractibility, we will
restrict the discussion to the particular forms $\phi(u) = Au^\alpha$, where
$A > 0$, and $\alpha > 0$ are parameters.

The same method of proof shows that for $\alpha = 1$, as before, but using
general $A$, we have 
\beqn
g(x;A;\alpha = 1) = -V^\prime(x;A;\alpha = 1) = g\parens{\frac{x}{A^2};A = 1;\alpha = 1}
\eeqn

It follows that 
\beqn
V(x;A;\alpha = 1) = A^2 V\parens{\frac{x}{A^2};A = 1;M = \frac{M}{A^2}}
\eeqn

This is as expected; the original model used $f$ as both the drift and the
diffusion parameter because one could scale time to make the diffusion equal
to one in appropriate time units. However if one wants to {\it compare} models,
then the parameter $A$ must be retained. If one does this, one sees that 
\beqn
g(x;A;\alpha = 1) \rightarrow g(0;A = 1;\alpha = 1) \equiv 1
\eeqn

so that 
\beqn
\lim_{A \rightarrow \infty} V(x;A;\alpha = 1) = \int_x^M du = M - x
\eeqn

The conclusion is that if the investor has the choice of investing in a risky
market or instead to be conservative by saving his salary without investing,
then the conservative strategy is asymptotically (as $A \rightarrow \infty$)
superior even though the resulting time to retirement is $M - x$, which is the
maximum delay among all models since there is no advantage to investment. The
conservative investment advice to avoid risk is usually given to older
investors; our modelling assumptions conclusions bear this out, even for young
investors in the limit as risk gets very large.

We next consider the case $A = 1$, and $\frac{1}{2} \le \alpha <1$. Since the
diffusion speed is larger for $\alpha < 1$ than for $\alpha  = 1$, one would
think that as $\alpha$ decreases the expected time would increase and investing
would be disadvantageous, but this is surprisingly not the case, as we see
below. Moreover we will show that for $A = 1$, and $0 < \alpha < \frac{1}{2}$,
one can find investment strategies that allow retirement in time $\epsilon$,
arbitrarily small. Another surprise is that there is a sharp discontinuity in
$V(x,\alpha)$ as $\alpha \uparrow \frac{1}{2}$. We show that $V(x,\alpha) = 0$
for $\alpha < \frac{1}{2}$ but as we see below
\beqn
V\parens{x;\alpha = \frac{1}{2}} = \frac{1}{2} \parens{e^{-2x} - e^{-2M}}
\eeqn

We turn to the $A > 0,\alpha \in [.5,1)$ problem: again we define 
$V^f_M(x) = \expesubsup{x}{f}{\tau_M}$, and we want to find $V_M(x) = \inf_f V^f_M(x)$.

Again we seek any function ${\bar{V}}(x)$ for which
\beqn
Y(t) = {\bar{V}}\parens{X^f(t)}+t
\eeqn

is a submartingale for any choice of $f$. Any such ${\bar{V}}(x)$ will be a
lower bound on $V(x)$ since we will again have
\beqn
\expesub{x}{{\bar{V}}(X^f(\tau_M))}+\expe{\tau^f_M} \ge \expesub{x}{{\bar{V}}(X^f(0))} = {\bar{V}}(x)
\eeqn

To be a submartingale, we need in the same way as in the case, $\alpha = 1$,
\beqn
{\bar{V}}^\prime(x)(1+f)+\frac{A^2}{2}{\bar{V}}^{\prime \prime}(x)f^{2\alpha} + 1 \ge 0, ~~0 \le x \le M, ~~f \ge 0
\eeqn

For fixed $x$, this is a minimum in $f$ at the point 
\beqn
f(x) = f_0(x) = \tothepow{\frac{-{\bar{V}}^\prime(x) }{{\bar{V}}^{\prime \prime}(x)}\frac{1}{A^2\alpha}}{\frac{1}{2\alpha -1}}
\eeqn

Requiring that $Y$ be a martingale for the best choice of $f = f_0$ above
gives an ode for $g(x) = g(x,A,\alpha) = -{\bar{V}}^\prime(x,A,\alpha)$.
After a calculation, very similar to the one above for $A = \alpha = 1$, the
ode is:
\beqn
\frac{-g^\prime(x)(1-g(x))^{2\alpha-1}}{g^{2\alpha}(x)} = \frac{\tothepow{1-\frac{1}{2\alpha}}{2 \alpha -1}}{A^2 \alpha}
\eeqn

Integrating gives
\beqn
\int_{g(x)}^1 \frac{(1-u)^{2\alpha -1}}{u^{2\alpha} } du = x \frac{\tothepow{1- \frac{1}{2\alpha}}{2 \alpha -1}}{A^2 \alpha} + c
\eeqn

Again we guess that $c = 0$ and we can solve for $g(x) \in [0,1]$ for any 
$x \ge 0$. We see that so long as $\alpha > .5$, there is no trouble. We can
write (since $V(M) = 0$),
\beqn
V(x) =\int_x^M -V^\prime(u) du = \int_x^M g(u)du &=& \int_x^M \frac{g(u)}{g^\prime(u)} g^\prime(u) du\\&=& b(\alpha) \int_0^{g(x)} \tothepow{\frac{1-u}{u}}{2\alpha -1}du
\eeqn

where 
\beqn
b(\alpha) = \frac{\tothepow{1-\frac{1}{2\alpha}}{2\alpha -1}}{\alpha A^2}
\eeqn

Finally we have that $f_0 =f_0(x,A,\alpha)$ is given by
\beqn
f_0(x) = \parens{\frac{1}{g(x)} - 1}\frac{1}{1-\frac{1}{2\alpha}}
\eeqn

\section{General Model in the Case of $\alpha < \frac{1}{2}$}\label{sec:alpha_lessthan_half}

The case $\alpha < \frac{1}{2}$ is especially interesting. The argument given
for guessing the optimal ${\hat{V}}$ breaks down.

\begin{thm}
There is no ${\hat{V}}$ that will make ${\hat{V}}(X^f(t) ) + t$ a submartingale for all
choices of $f = f(x)$ except the trivial case ${\hat{V}}(x) \equiv 0$. This is
the best lower bound that submartingale theory can provide which lead one to 
suspect that $V(x;\alpha) \equiv 0$ for $\alpha < \frac{1}{2}$. \end{thm}

\begin{proof} Here we assume that $A=1$. We need to find an $f$ that makes the expected time to reach $M$ arbitrarily
small. Consider the investment strategy
\beqn
f(x) = 0, 0 \le x \le \epsilon, ~~f(x) = c, ~~\epsilon < x < M
\eeqn

If we can find a function, $g = g(x) = g_{\epsilon,c}(x), 0 \le x \le M$ for
which $Y(t) = g(X_t) + t$ is a martingale and $g(M) = 0$, then the martingale
theorem gives that $g(x) = Y(0) = \expesub{x}{Y(\tau_M)} = \expesub{x}{\tau_M}$. Ito calculus
gives that $g$ must be, for appropriate constants, $A^\prime,B,D$,
\beqn 
g(x) = \begin{cases}
D - x, ~~&0 \le x \le \epsilon \\
\frac{M-x}{1+c} + A^\prime\parens{\exp{-\frac{2(1+c)}{c^{2\alpha} }x} - \exp{-\frac{2(1+c)}{c^{2\alpha}}M}}, &\epsilon \le x \le M
\end{cases}
\eeqn

We determine from continuity at $x = \epsilon$ of $g$ and $g^\prime$ that
\beqn
A^\prime  &=& \frac{c^{2\alpha + 1}}{2(1+c)^2}\,\exp{\frac{2(1+c)}{c^{2\alpha}} \epsilon}\\
D &=& \frac{M-\epsilon}{1+c} + A^\prime\parens{\exp{-\frac{2(1+c)}{c^{2\alpha}}\epsilon} - \exp{-\frac{2(1+c)}{c^{2\alpha} }x}}, ~~\epsilon \le x \le M
\eeqn

Finally, set $\epsilon = \frac{1}{1+c}$, and let $c \rightarrow \infty$ and
verify that $A^\prime$ and $D$ tend to zero as $c \rightarrow \infty$ and we
have shown that $V(x;\alpha) \equiv 0$ for $\alpha < \frac{1}{2}$. 
\end{proof}

\begin{remark}

Under this model one can retire in arbitrarliy small expected time. This should probably
be interpreted that the model with $\alpha < \frac{1}{2}$ does not represent
the real world; models should be looked at carefully and rejected if they do
not conform to reality, unless of course, reality is incorrect.

\end{remark}

\acks
We thank Professor Lawrence Brown and Professor Michael Harrison for their invaluable advice.


%
%
%

\bibliographystyle{apt}
\bibliography{refs}

\begin{thebibliography}{1}

\bibitem{Scholes}
{\sc Black, F. and Scholes, M.} (1973).
\newblock The pricing of options and corporate liabilites.
\newblock {\em Journal of Political Economy\/}.

\bibitem{Cover}
{\sc Cover, T.~M.} (1991).
\newblock Universal portfolios.
\newblock {\em Mathematical Finance\/}.

\bibitem{Foster}
{\sc Foster, D~Kakade, S. and Ronen, O.}
\newblock Early retirement using leveraged investments.
\newblock 2007.

\bibitem{IandM}
{\sc Ito, K. and McKean, H.~P.} (1965).
\newblock {\em Diffusion Processes and Their Sample Paths}.
\newblock Springer.

\bibitem{KandS}
{\sc Karatzas, I. and Shreve, S.} (1991).
\newblock {\em Brownian Motion and Stochastic Calculus}.
\newblock Springer-Verlag.

\bibitem{KandT}
{\sc Karlin, S. and Taylor, H.~M.} (1981).
\newblock {\em A Second Course in Stochastic Processes}.
\newblock Academic Press.

\bibitem{HPM}
{\sc McKean, H.~P.} (1969).
\newblock {\em Stochastic Integrals}.
\newblock Academic Press.

\bibitem{Merton2}
{\sc Merton, R.~C.} (1969).
\newblock Lifetime portfolio selection under uncertainty: The continuous-time
  case.
\newblock {\em Rev. Economics Statist.\/}.

\bibitem{Merton}
{\sc Merton, R.~C.} (1992).
\newblock {\em Continuous-Time Finance}.
\newblock Blackwell Publishers.

\end{thebibliography}

\end{document}